\documentclass[a4paper,11pt]{article}
\pdfoutput=1 

\usepackage{jheppub} 

\usepackage[T1]{fontenc} 
\usepackage{graphicx}
\usepackage{amsmath}
\usepackage{amssymb}
\usepackage[dvipsnames]{xcolor}
\usepackage{centernot}

\usepackage{cancel}

\def\U1mt{U(1)_{L_\mu-L_\tau}}
\newcommand{\la}{\lambda}

\newcommand{\ga}{\gamma}
\newcommand{\si}{\sigma}
\newcommand{\De}{\Delta}
\newcommand{\ol}{\overline}
\newcommand{\wt}{\widetilde}
\newcommand{\nl}{\nonumber\\}

\newcommand{\eq}[1]{\begin{align}#1\end{align}}

\title{\boldmath A connection between flavour anomaly, neutrino mass,  and axion}

\author{Seungwon Baek,}
\emailAdd{sbaek@korea.ac.kr}
\affiliation{Department of Physics, Korea University, \\
Anam-ro 145, Sungbuk-gu, Seoul 02841, Korea}

\abstract{We propose a minimal model in which the  flavour anomaly in the $b \to s \mu^+ \mu^-$ transition
is connected to the breaking of Peccei-Quinn (PQ) symmetry. The
flavour anomaly is explained from new physics contribution by introducing one generation of heavy quark and heavy lepton which are vector-like under 
the standard model (SM) gauge group but charged under a local $U(1)_X$ group.
They mix with the SM quarks and leptons, inducing flavour-changing $Z^\prime$ couplings, which generates the $b \to s \mu^+\mu^-$ anomaly at tree level.
On the other hand the new fermions are chiral under the global Peccei-Quinn(PQ)  symmetry.
The pseudo-Goldstone boson coming from the spontaneous breaking of the
PQ symmetry becomes an axion, solving the strong CP problem and providing a cold dark matter candidate.
The same symmetry prevents the right-handed neutrino from having a Majorana mass term.
But the introduction of a neutrino-specific Higgs doublet allows neutrino to have Dirac mass term without fine-tuning problem.
The model shows an interplay between axion, neutrino, dark matter, and flavour physics.}

\begin{document} 
\maketitle
\flushbottom

\section{Introduction}
\label{sec:intro}

Although the standard model (SM) has passed the experimental tests successfully for decades, there are hints that suggest more fundamental 
theory beyond the SM. These include the existence of dark matter (DM), the neutrino mass and mixing, and the strong CP problem.
In addition, there are some tantalizing anomalies in the $B$ decay data from the Belle and the LHCb experiments, 
which may also require new physics (NP) beyond the SM.
In this paper we consider a NP model which addresses the above problems of the SM simultaneously.

The most popular solution of the strong CP problem is the introduction of axion which is a
pseudo-Goldstone boson coming from the breaking of the global Peccei-Quinn (PQ) symmetry, $U(1)_{\rm PQ}$.
In the paper \cite{Baek:2019wdn} we suggested a NP model where the neutrino mass is generated from the breaking of PQ symmetry, thereby the explanation of the neutrino masses
and axion can be unified.  In the model two Higgs doublets are introduced, the SM-like Higgs doublet ($\Phi_2$) couples to the quarks and charged leptons, whereas
the new Higgs doublet ($\Phi_1$) couples solely to the neutrino sector. In this neutrino-specific two Higgs doublet model ($\nu$THDM), the Higgs $\Phi_1$ and
the right-handed neutrinos $\nu_{i_R}$ $(i=1,2,3)$ as well as $S$ whose phase is the main component of the axion are charged under the PQ symmetry~\cite{Baek:2016wml,Baek:2018wuo}. 
The PQ symmetry prohibits the right-handed neutrino from having mass term, making the type-I seesaw mechanism not effective.
But it allows the light neutrino to have Dirac-type mass without fine-tuning problem. There is a simple seesaw-like relation for the vacuum expectation
value (VEV) of $\Phi_1$:
\eq{
 v_1 \approx \frac{\mu v_2 v_S}{M_{\Phi_1}^2},
}
where $v_2 \equiv \sqrt{2} \langle \Phi_2^0 \rangle$, $v_S \equiv \sqrt{2} \langle S \rangle$, $M_{\Phi_1}$ is the mass scale of $\Phi_1$, and
$\mu$ is the coupling constant of trilinear interaction, $ \mu \Phi_1^\dagger \Phi_2 S$.
For $M_{\Phi_1} \sim v_S \sim 10^{12}$ GeV, $v_2 \simeq 246$ GeV, and $\mu \sim 1$ GeV, we get $v_1 \sim 0.1$ eV, which renders the neutrino
mass, $m_\nu (=y_\nu v_1/\sqrt{2})$, at sub-eV scale when neutrino Yukawa coupling is of order one, $y_\nu \sim {\cal O}(1)$.
By introducing vector-like heavy quarks or additional Higgs doublets, we can introduce either KSVZ-type or DFSZ-type axions~\cite{Baek:2019wdn}.

The flavour-changing neutral current (FCNC) processes are  very sensitive probes of NP~\cite{Buchalla:1995vs,Baek:1998yn,Baek:2000sj,Baek:2002rt}. 
Especially the quark-level process, $b \to s \mu^+ \mu^-$, has been drawing much interest during recent years due to discrepancies between the experimental measurements
and the theoretical predictions.
A measurement of particular interest is the ratio of branching fractions,
\eq{
R_{K^{(*)}}[q_{\rm min}^2 < q^2 < q_{\rm max}^2] =\frac{\int_{q^2_{\rm min}}^{q^2_{\rm max}} \frac{d\Gamma[B \to K^{(*)} \mu^+ \mu^-]}{dq^2} dq^2}
{\int_{q^2_{\rm min}}^{q^2_{\rm max}} \frac{d\Gamma[B \to K^{(*)} e^+ e^-]}{dq^2} dq^2},
}
where $q^2$ is the dilepton mass squared. In the SM the gauge interactions of all three charged leptons are identical, and the ratio is predicted to be
$1$ up to small corrections related to the lepton mass.
The experimental reports are
\eq{
R_K[1.1 \, {\rm GeV^2}< q^2 < 6.0 \, {\rm GeV^2}] &= 0.846^{+0.060+0.016}_{-0.054-0.014} \; (\text{LHCb~\cite{Aaij:2019wad}}),\nl
R_{K^*}[1.1 \, {\rm GeV^2}< q^2 < 6.0 \, {\rm GeV^2}] &= 0.69^{+0.11}_{-0.07}\pm 0.05, \; (\text{LHCb~\cite{Aaij:2017vbb}}), \nl
R_{K^*}[0.1 \, {\rm GeV^2}< q^2 < 8.0 \, {\rm GeV^2}] &= 0.90^{+0.27}_{-0.21}\pm 0.10, \; (\text{Belle~\cite{Abdesselam:2019wac}}),
}
which show violation of the lepton flavour universality (LFU).
Combining with other $b \to s \ell^+ \ell^-$ and $b \to s \gamma$ observables, the global fits show the SM is disfavoured with large
significancies~\cite{Alguero:2019ptt,Alok:2019ufo,Ciuchini:2019usw,Datta:2019zca,Aebischer:2019mlg,Kowalska:2019ley,Arbey:2019duh}.
For example, for the scenario $C_9^{\mu, {\rm NP}}=-C_{10}^{\mu, {\rm NP}}$ which we will take in this paper, 
the pull with respect to the SM
is $5.2 \sigma$ with the best fit value~\cite{Alguero:2019ptt},
\eq{
C_9^{\mu, {\rm NP}}=-C_{10}^{\mu, {\rm NP}}=-0.46.
\label{eq:C9_NP}
}
The 1$\sigma$ and 2$\sigma$ regions are $[-0.56,-0.37]$ and $[-0.66,-0.28]$, respectively~\cite{Alguero:2019ptt}.

Here the effective Hamiltonian for the $C_{9(10)}$ is defined as
\eq{
{\cal H}_{\rm eff} = -\frac{4 G_F}{\sqrt{2}}  \, \frac{\alpha_{\rm em}}{4 \pi} \, V_{ts}^* V_{tb}
\sum_{\ell=e,\mu,\tau} \ol{s}_L \ga_\mu b_L\;(C_9^\ell \,\ol{\ell} \ga^\mu \ell + C_{10}^\ell \, \ol{\ell} \ga^\mu \ga_5 \ell ).
}
The anomaly can be explained in numerous NP models. Among them, many 
models~\cite{Sierra:2015fma,Arnan:2016cpy,Cline:2017lvv,Kawamura:2017ecz,
Baek:2017sew,Cline:2017aed,Chiang:2017zkh,Cline:2017qqu,Vicente:2018xbv,
Falkowski:2018dsl,Baek:2018aru,Darme:2018hqg,Barman:2018jhz,Singirala:2018mio,
Vicente:2018frk,Baek:2019qte,Cerdeno:2019vpd,Ko:2019tts,Biswas:2019twf,Arnan:2019uhr,Trifinopoulos:2019lyo,
Han:2019diw,Darme:2020hpo} show an interesting interplay between the flavour physics and dark matter
in which the WIMP cold DM (CDM) is related to the mechanism explaining the 
flavour anomaly or contributes to the $b \to s \mu^+ \mu^-$ process through loop. 

In this paper we present a NP model which shows connection between the $b \to s \mu^+\mu^-$ anomaly, the axion, and the neutrino mass.
Since the axion is a good candidate for CDM, the model can address the dark matter candidate, the strong CP problem, the neutrino mass,
and the flavour problem at the same time. In the model super-heavy scalar $S$, vector-like quark doublets $Q_{L,R}$, and lepton doublets $L_{L,R}$
are introduced. They are charged under the global $U(1)_{\rm PQ}$ symmetry. Thereby the pseudo-scalar component of $S$ becomes an axion
of the KSVZ-type~\cite{Kim:1979if,Shifman:1979if}.

The vector-like fermions are also charged under a local $U(1)_X$ symmetry. The SM fields are neutral under $U(1)_X$.
The gauge boson $Z'$ of the $U(1)_X$ can still interact with the SM fermions through the mixing with the heavy vector-like fermions
after the gauge symmetry is spontaneously broken by the VEV of $U(1)_X$-charged scalar $\phi$. This allows tree-level diagrams to produce
FCNC processes such as $b \to s \ell^+ \ell^-$ and $B_s-\ol{B}_s$ mixing.

The paper is organised as follows:
we introduce our model in the next section.
In Section~\ref{sec:axion} we investigate the axion properties in our model and compare with those in the original KSVZ model.
We also briefly review the neutrino mass generation.
In Section~\ref{sec:flavour} we resolve the flavour problem by introducing the NP $b \to s\mu^+\mu^-$ transition at tree-level.
We discuss the constraints.
In Section~\ref{sec:conclusion} we conclude the paper.

\section{The Model}
\label{sec:model}

The new particles in the model as well as the SM ones are shown in Table~\ref{tab:1} with their representations under the SM gauge group
$SU(3)_C \times SU(2)_L \times U(1)_Y$ and their charges under the local $U(1)_X$ symmetry and the global $U(1)_{\rm PQ}$ symmetry.

\begin{table}[ht]
\renewcommand{\arraystretch}{1.5}
\begin{center} 
\begin{tabular}{|c||c|c|c|c||c|c|c|c|c|c||c|c|c|c|}\hline\hline  
&\multicolumn{4}{c||}{Scalars} & \multicolumn{10}{c|}{Fermions} \\\hline
  & $\Phi_1$          & $\Phi_2$             & $\phi$               & $S$ 
  & $q_{i_L}$           & $u_{i_R}$             & $d_{i_R}$           
  & $\ell_{i_L}$        & $e_{i_R}$             & $\nu_{i_R}$ 
  & $ Q_L$            & $Q_R$         & $ L_L$            & $L_R$    \\
  \hline 
$SU(3)_C$ 
  & $\boldsymbol{1}$         & $\boldsymbol{1}$             & $\boldsymbol{1}$             & $\boldsymbol{1}$
  & $\boldsymbol{3}$         & $\boldsymbol{3}$             & $\boldsymbol{3}$             
  & $\boldsymbol{1}$         & $\boldsymbol{1}$             & $\boldsymbol{1}$ 
  & $\boldsymbol{3}$         & $\boldsymbol{3}$        & $\boldsymbol{1}$         & $\boldsymbol{1}$ \\
\hline
$SU(2)_L$ 
  & $\boldsymbol{2}$         & $\boldsymbol{2}$             & $\boldsymbol{1}$             & $\boldsymbol{1}$
  & $\boldsymbol{2}$         & $\boldsymbol{1}$             & $\boldsymbol{1}$             
  & $\boldsymbol{2}$         & $\boldsymbol{1}$             & $\boldsymbol{1}$ 
  & $\boldsymbol{2}$         & $\boldsymbol{2}$        & $\boldsymbol{2}$         & $\boldsymbol{2}$ \\ 
\hline
$U(1)_Y$ 
  & $\frac12$        & $\frac12$             & $0$             & $0$
  & $\frac16$        & $\frac23$             & $-\frac13$      
  & $-\frac12$       & $-1$                     & $0$        
  & $\frac16$        & $\frac16$          & $-\frac12$        & $-\frac12$  \\
\hline
$U(1)_X$ 
  & $0$                 & $0$                      & $1$             & $0$
  & $0$                 & $0$                      & $0$      
  & $0$                 & $0$                      & $0$        
  & $-1$                & $-1$             & $-1$                & $-1$ \\
\hline
$U(1)_{\rm PQ}$ 
  & $X_1$                 & $X_2$                      & $X_\phi$             & $X_S$
  & $X_{q_L}$                 & $X_{u_R}$                      & $X_{d_R}$      
  & $X_{\ell_L}$                 & $X_{e_R}$                      & $X_{\nu_R}$        
  & $X_{Q_L}$                 & $X_{Q_R}$          & $X_{L_L}$                 & $X_{L_R}$  \\
\hline 
\end{tabular}
\end{center}
\caption{The particles and their charges under $SU(3)_C \times SU(2)_L \times U(1)_Y \times U(1)_X \times U(1)_{\rm PQ}$ in our model.  
The PQ charges, the $X$'s, are real numbers, satisfying (\ref{eq:X}) in the text.}
\label{tab:1}
\end{table}

The scalar potential is in the form
\eq{
V &= \mu_1^2 \Phi_1^\dagger \Phi_1 -\mu_2^2 \Phi_2^\dagger \Phi_2 -\mu_S^2 S^* S -\mu_\phi^2 \phi^* \phi
       -(\mu \Phi_1^\dagger \Phi_2 S + h.c.) \nl
   &+ \lambda_1 (\Phi_1^\dagger \Phi_1)^2 + \lambda_2 (\Phi_2^\dagger \Phi_2)^2
+ \lambda_3 (\Phi_1^\dagger \Phi_1) (\Phi_2^\dagger \Phi_2) + \lambda_4 (\Phi_1^\dagger \Phi_2) (\Phi_2^\dagger \Phi_1) \nl
&+ \lambda_\phi (\phi^* \phi)^2 +\lambda_{1\phi} (\Phi_1^\dagger \Phi_1) (\phi^* \phi)+\lambda_{2\phi} (\Phi_2^\dagger \Phi_2) (\phi^* \phi) \nl
&+ \lambda_S (S^* S)^2 +\lambda_{1S} (\Phi_1^\dagger \Phi_1) (S^* S)+\lambda_{2S} (\Phi_2^\dagger \Phi_2) (S^* S)+\lambda_{\phi S} (\phi^* \phi)(S^* S).
\label{eq:V}
}
The scalar potential contains the $\mu$-term, $(-\mu \Phi_1^\dagger \Phi_2 S + h.c.)$, which plays an essential role in the generation
of the neutrino mass~\cite{Baek:2019wdn}. To keep the hierarchy between the PQ scale ($\sim 10^{12}$ GeV), the $U(1)_X$ scale ($\sim 10^5$ GeV),
the EW scale ($\sim 100$ GeV), and the neutrino mass scale ($\sim 0.1$ eV) from the radiative corrections, we need to suppress the
corresponding mixing parameters $\la_{\phi S}$, $\la_{1(2) S}$,  $\la_{1(2) \phi}$. The smallness of these parameters is technically natural
due to the extended Poincar\'e symmetry~\cite{Foot:2013hna,Baek:2019wdn}. 

The Yukawa interactions are
\begin{align}
{\cal L}_{\Phi} &=
 -y^u_{ij} \ol{q}_{i_L} \wt{\Phi}_2 u_{j_R} 
    -y^d_{ij} \ol{q}_{i_L} \Phi_2 d_{j_R} 
    -y^e_{ij} \ol{\ell}_{i_L} \Phi_2 e_{j_R} 
    -y^\nu_{ij} \ol{\ell}_{i_L} \wt{\Phi}_1 \nu_{j_R} +h.c. 
\label{eq:Yukawa_Phi}    \\
{\cal L}_{\phi} &=    
 -\la^q_i \ol{q}_{i_L} Q_R \phi -\la^\ell_i \ol{\ell}_{i_L} L_R \phi + h.c. 
\label{eq:Yukawa_phi}    \\
{\cal L}_{S} &=    
 -y_Q \ol{Q}_L Q_R S  -y_L \ol{L}_L L_R S  + h.c., 
\label{eq:Yukawa_S}    
\end{align}
where $i,j=1,2,3$ are generation indices, and $\wt{\Phi}_{1(2)} = i \sigma^2 \Phi^*_{1(2)}$. 
We assign the PQ charges in such a way that only the terms in
the interactions (\ref{eq:V}), (\ref{eq:Yukawa_Phi}), (\ref{eq:Yukawa_phi}), and (\ref{eq:Yukawa_S}) 
are allowed. The interactions are already invariant under the SM gauge group and $U(1)_X$.
We notice that 
all the quarks and charged-leptons are coupled to the Higgs doublet $\Phi_2$, whereas the neutrinos are
coupled to the Higgs doublet $\Phi_1$. 
It is also noted that the right-handed neutrinos do not have the
Majorana mass terms because they are charged under $U(1)_{\rm PQ}$.
The $Q$'s and $L$'s are vector-like under both the SM gauge and $U(1)_X$ gauge symmetries.
Therefore gauge anomalies cancel in our model.

\section{The Axion and the Dirac neutrino}
\label{sec:axion}

To explain the flavour anomaly in $b \to s \mu^+\mu^-$ transition, we need to fix the properties of the heavy vector-like quarks and leptons. 
A minimal model for the scenario (\ref{eq:C9_NP}) is to introduce $SU(2)_L$-doublet vector-like quarks and leptons, $Q_{L,R}$ and $L_{L,R}$:
\eq{
Q_{L,R} = 
\begin{pmatrix}
U_{L,R} \\
D_{L,R} \\
\end{pmatrix}, \quad
L_{L,R} = 
\begin{pmatrix}
N_{L,R} \\
E_{L,R} \\
\end{pmatrix}. 
}
As a consequence they have a definite model-dependent predictions for the axion properties, such as axion couplings to 
the photons and to the other SM particles.

We first fix the PQ charges of the scalar fields.
We can identify the axion
by defining~\cite{Peccei:2006as}\footnote{The axion field can be identified
also by explicit orthogonalization~\cite{Srednicki:1985xd}.},
\eq{
\Phi_1 ={v_1 \over \sqrt{2}} e^{i X_1 a/v_{\rm PQ}} 
\begin{pmatrix}
0 \\ 1
\end{pmatrix}, \quad
\Phi_2 ={v_2 \over \sqrt{2}} e^{i X_2 a/v_{\rm PQ}} 
\begin{pmatrix}
0 \\ 1
\end{pmatrix}, \quad
\phi = {v_\phi \over \sqrt{2}} e^{i X_\phi a/v_{\rm PQ}}, \quad
S = {v_S \over \sqrt{2}} e^{i X_S a/v_{\rm PQ}}, 
\label{eq:nonlinear}
}
where $a \to a + \alpha v_{\rm PQ}$ under $U(1)_{\rm PQ}$ transformation.
We can determine the $X$'s as in~\cite{Baek:2019wdn}:
\eq{
X_1 = \frac{v_2^2}{v_1^2+v_2^2}, \quad
X_2 =- \frac{v_1^2}{v_1^2+v_2^2}, \quad
X_\phi = 0, \quad
}
where we take the nomalization, $X_S =1$. They satisfy $-X_1+X_2+X_S=0$ as required by the trilinear term in (\ref{eq:V}).
The canonical normalization for the kinetic
energy term of $a$ is obtained with
\eq{
v_{\rm PQ} = \left( \sum_{i=1,2,\phi,S} X_i^2 v_i^2\right)^{1/2}.
}
As can be seeen from (\ref{eq:Yukawa_Phi}), (\ref{eq:Yukawa_phi}), and (\ref{eq:Yukawa_S}), the other PQ-charges are related as follows:
\eq{
 -X_{q_L}-X_2+X_{u_R} =0, &\quad -X_{q_L}+X_2+X_{d_R} =0, \nl
  -X_{\ell_L}+X_2+X_{e_R} =0, &\quad -X_{\ell_L}-X_1+X_{\nu_R} =0, \nl
 -X_{q_L} + X_{Q_R} =0, &\quad -X_{\ell_L} + X_{L_R} =0, \nl
 -X_{Q_L} + X_{Q_R} +1=0, &\quad -X_{L_L} + X_{L_R} +1=0.
 \label{eq:X}
}


By expanding (\ref{eq:nonlinear}) we can also write the axion $a$ as 
\eq{
a &= {1 \over v_{\rm PQ}} \sum_{i=1,2,\phi,S} X_i v_i a_i,
}
where $a_i$'s ($i=1, 2, \phi, S$) are the pseudo-scalar components of scalar fields $\Phi_1, \Phi_2, \phi, S$ whose VEVs are $v_i$. 

The PQ-current
\eq{
J_{\rm PQ}^\mu &= v_{\rm PQ} \partial^\mu a -{X_2 \over 2} \sum_i \ol{u}_i \ga^\mu \ga_5 u_i +{X_2 \over 2} \sum_i \ol{d}_i \ga^\mu \ga_5 d_i
+{X_2 \over 2} \sum_i \ol{e}_i \ga^\mu \ga_5 e_i \nl
&-{X_1 \over 2} \sum_i \ol{\nu}_i \ga^\mu \ga_5 \nu_i
+{1 \over 2} \ol{Q} \ga^\mu \ga_5 Q
+{1 \over 2} \ol{L} \ga^\mu \ga_5 L,
}
satisfies anomaly equation:
\eq{
\partial_\mu J_{\rm PQ}^\mu = -{N g_s^2 \over 16 \pi^2} G^a_{\mu\nu} \wt{G}^{a\mu\nu}
-{E e^2 \over 16 \pi^2} F_{\mu\nu} \wt{F}^{\mu\nu},
}
where $G^a_{\mu\nu} (F_{\mu\nu})$ is the gluon (electromagnetic) field strength tensor and $\wt{G}^a_{\mu\nu} (\wt{F}_{\mu\nu})$ is its dual tensor.
We obtain 
\eq{
N = 1,\quad E= {8 \over 3}.
}
We note that the QCD and the electromagnetic anomalies from the SM fermions cancel and
the non-trivial contributions come entirely from the new heavy fermions.
In this model the axion domain wall number, given by
\eq{
N_{\rm DW} = 2 N = 2,
}
is different from that of the KSVZ  (DFSZ) model where $N_{\rm DW} = 1 (6)$~\cite{Kim:1979if,Shifman:1979if,Dine:1981rt,Zhitnitsky:1980tq}.

The axion mass is 
\eq{
m_a = \frac{f_\pi m_\pi}{f_a} \frac{\sqrt{z}}{1+z} \simeq 5.7 \,{\rm \mu eV}\left(10^{12} \, {\rm GeV} \over f_a\right),
}
where $m_\pi \simeq 135$ MeV is the neutral pion mass,
$f_\pi \simeq 92.3$ MeV is the pion decay constant, $f_a =v_{\rm PQ}/N_{\rm DW}$ is the axion decay constant, and $z=m_u/m_d \simeq 0.472$ is the light
quarks mass ratio.
The axion interaction with the photon can written in the form
\eq{
{\cal L}_{a\ga\ga} = -{g_{a\ga\ga} \over 4} F_{\mu\nu} \wt{F}^{\mu\nu} a = g_{a\ga\ga} \boldsymbol{E}\cdot \boldsymbol{B} a,
}
where
\eq{
g_{a\ga\ga} = \frac{\alpha_{\rm em}}{2 \pi f_a} \left( {E \over N}-{2 \over 3} \,\frac{4+z}{1+z} \right) \quad \text{with} \;{E \over N}={8 \over 3}.
\label{eq:arr}
}
The axion-photon coupling (\ref{eq:arr}) agrees with that of the DFSZ model {\em not} the KSVZ model.
The axion interaction with fermions can be written in the form,
\eq{
{\cal L}_{a f f } = \frac{C_f }{2 f_a} \partial_\mu a \ol{f} \ga^\mu \ga_5 f.
}
The tree-level axion coupling to electrons and neutrinos are
\eq{
C_e = {X_2 \over 2}, \quad C_\nu = -{X_1 \over 2},
}
which can be compared with the KSVZ model where $C_e = C_\nu=0$.
Since $X_2$ is tiny in our model,  the loop-induced $C_e$ is much larger~\cite{Srednicki:1985xd}.
However, the tree-level $C_\nu$ is of order unity, and may be probed at future neutrino oscillation experiments~\cite{Huang:2018cwo}.


%
\begin{figure}[htbp]
\begin{center}
\includegraphics[width=0.6\textwidth]{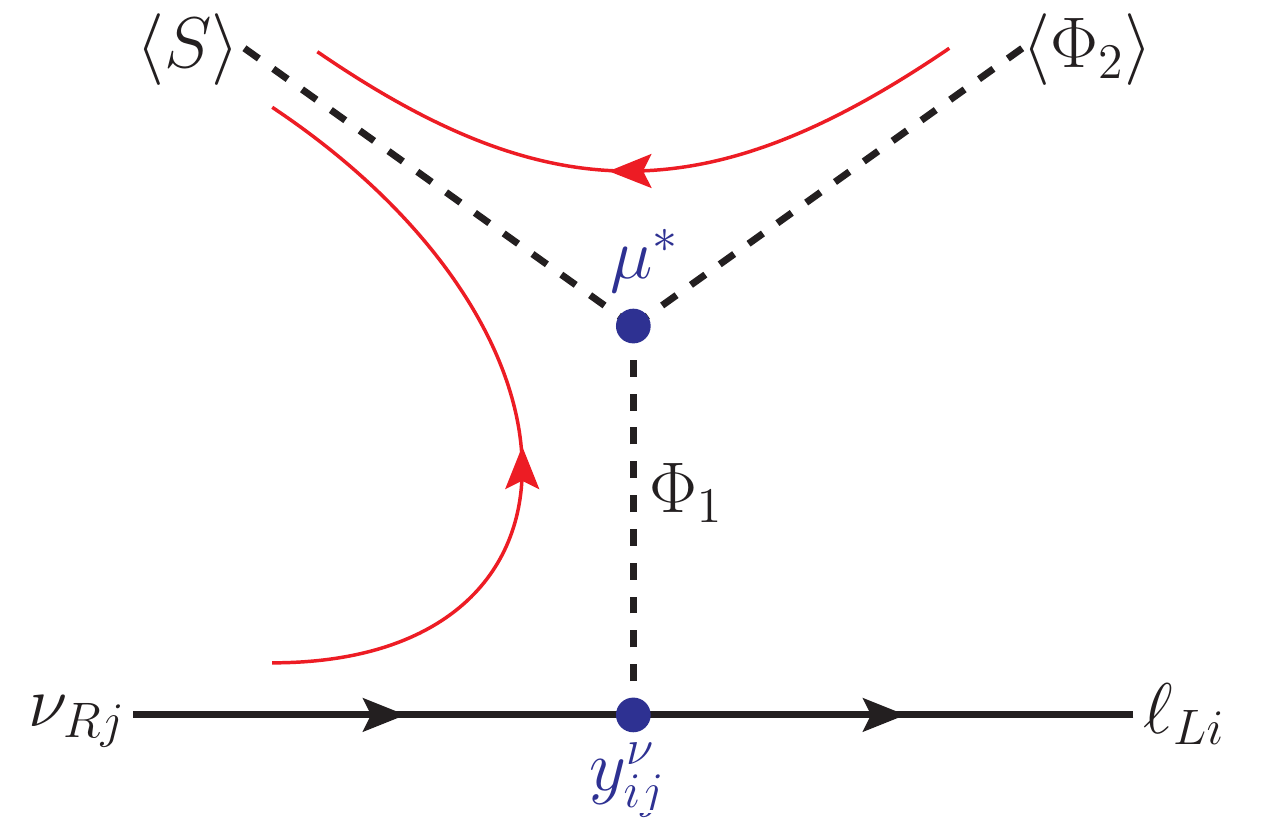}
\caption{Feynman diagram for the neutrino mass generation. The red (black) arrows represent the flow of 
{ $U(1)_{\rm PQ}$} (lepton number) current.}
\label{fig:neutrino_mass}
\end{center}
\end{figure}
Now we briefly review the mechanism for the neutrino mass generation suggested in~\cite{Baek:2019wdn}. 
The tree-level diagram shown in Fig.~\ref{fig:neutrino_mass} generates the neutrino mass.
The red (black) arrows represent the flow of { $U(1)_{\rm PQ}$} (lepton number) current when we set $X_{\ell_L}=0$.
The Yukawa interaction and the $\mu$-term generates the Dirac neutrino masses after the $S$ and $\Phi_2$ fields get
VEVs:
\eq{
m_{ij}^\nu = \frac{y_{ij}^\nu \mu v_2 v_S}{2 M^2_{\Phi_1}} \sim 0.1 \, y_{ij}^\nu \left(\mu \over 1 \, {\rm GeV}\right) 
\left(v_2 \over 246 \, {\rm GeV}\right)  \left(v_S \over 10^{12} \, {\rm GeV}\right) \left(10^{12} \, {\rm GeV} \over M_{\Phi_1}\right)^2 \, {\rm eV}.
\label{eq:m_nu}
}
The seesaw-like formula shows that the neutrino masses are ${\cal O}(0.1)$ eV for $y_{ij}^\nu \sim 1$, when  $\mu \sim 1$ GeV, 
$v_S \sim M_{\Phi_1} \sim 10^{12}$ GeV. 
Other studies on the link between $U(1)_{\rm PQ}$ symmetry and neutrino/flavour can be found 
in~\cite{Berezhiani:1989fp,Gu:2006dc,Chen:2012baa,Dasgupta:2013cwa,Bertolini:2014aia,Ahn:2015pia,Gu:2016hxh,Ma:2017zyb,Suematsu:2017kcu,Ahn:2018cau,
Reig:2018yfd,Carvajal:2018ohk,Ahn:2019add,delaVega:2020jcp,CentellesChulia:2020bnf}.

\section{The $b \to s \mu^+\mu^-$ transition}
\label{sec:flavour}

The new colored fermions $Q_{L,R}$ induce the QCD anomaly, making $a$ an axion candidate to solve the strong CP problem. 
The heavy quarks (heavy leptons) can also mix with the SM quarks (leptons).
The mixing can  generate $Z^\prime-b-s$ ($Z^\prime-\mu-\mu$) vertex at tree-level, which can induce $b \to s \mu^+ \mu^-$ 
transition at tree-level. 

In Table~\ref{tab:1} we introduced 
heavy vector-like quarks $Q_{L,R}$ and leptons $L_{L,R}$ in the same representation under the SM gauge group with
$q_L$ and $\ell_L$, respectively.
The heavy vector-like fermions $Q_L (L_L)$ can mix with the left-handed quarks (leptons) $q_{i_L} (\ell_{i_L})$ when
$\phi$ gets VEV, $v_\phi$.
Since the $Z^\prime$ couples only to the left-handed quarks and leptons, we can achieve the scenario in (\ref{eq:C9_NP}).


In the CKM basis where the SM Yukawa couplings are diagonal, we can write (\ref{eq:Yukawa_phi}) as
\eq{
{\cal L} = -( \ol{u}_{i_L} V^{\rm CKM}_{ij} \la^q_j U_R + \ol{d}_{i_L} \la^q_i D_R) \phi + h.c.,
}
where we assumed the quark mixing represented by the CKM matrix $V_{\rm CKM}$ arises only in the up-quark sector. 
When we include the new heavy quark, the down-type quark mass matrix in the CKM basis has off-diagonal components,
\eq{
{\cal L}_{\rm mass} =  - (\ol{d}_L, \ol{s}_L, \ol{b}_L, \ol{D}_L)
\begin{pmatrix}
m_d &  0          &   0        &      {1 \over \sqrt{2}} \lambda_d v_\phi  \\
 0     &  m_s     &   0         &      {1 \over \sqrt{2}} \lambda_s v_\phi  \\
 0     &  0          &   m_b    &      {1 \over \sqrt{2}} \lambda_b v_\phi  \\
  0    &  0          &   0         &      M_Q  \\
\end{pmatrix}
\begin{pmatrix}
d_R \\
s_R \\
b_R \\
D_R
\end{pmatrix}+h.c.,
\label{eq:d_matrix}
}
where $M_Q \equiv M_D =M_U=y_Q v_S/\sqrt{2}$, and $(\lambda_d, \lambda_s, \lambda_b) \equiv (\lambda^q_1, \lambda^q_2, \lambda^q_3)$.
The corresponding up-quark mass matrix is in the same form with replacement $(m_d, m_s, m_b)$ and $\lambda^q_i$ by
$(m_u, m_c, m_t)$ and $(V_{\rm CKM} \lambda^q)_i$, respectively. The mass matrix (\ref{eq:d_matrix}) can be
diagonalized by biunitary transformation:
\eq{
{\cal V}_L^{d \dagger}  \, {\cal M}_d \, {\cal V}_R^d \simeq {\rm diag}(m_d, m_s, m_b, M_Q),
\label{eq:biunitary}
}
where ${\cal M}_d \equiv$ (4 $\times$ 4 mass matrix) in (\ref{eq:d_matrix}). 
We obtain approximately
\eq{
{\cal V}_R^d &\simeq {\rm diag}(1,1,1,1), \nl
{\cal V}_L^d &\simeq 
\begin{pmatrix}
1      &  0          &   0        &      \frac{\lambda_d v_\phi}{\sqrt{2} M_Q}  \\
 0     & 1           &   0         &     \frac{\lambda_s v_\phi}{\sqrt{2} M_Q}  \\
 0     &  0          &   1         &     \frac{\lambda_b v_\phi}{\sqrt{2} M_Q}   \\
 - \frac{\lambda_d^* v_\phi}{\sqrt{2} M_Q}     &  - \frac{\lambda_s^* v_\phi}{\sqrt{2} M_Q}        &    -\frac{\lambda_b^* v_\phi}{\sqrt{2} M_Q}        &      1 \\
\end{pmatrix}.
\label{eq:V_mix}
}
The relations (\ref{eq:biunitary}) and (\ref{eq:V_mix}) are correct up to ${\cal O}(\lambda_{\rm max} v_\phi / \sqrt{2} M_Q)^2$ 
with $\lambda_{\rm max} = \text{max}(\la_d,\la_s,\la_b)$.

Then the effective $Z'-s-b$ vertex for $b \to s$ transition is
\eq{
\Lambda_L^{sb} \, Z^{\prime \alpha}\, \ol{s}_L \gamma_\alpha b_L + h.c.
\equiv g_X \left({\cal V}_L^d\right)^*_{42} \left({\cal V}_L^d\right)_{43}\, Z^{\prime \alpha}\, \ol{s}_L \gamma_\alpha b_L + h.c.,
\label{eq:Zprime-s-b}
}
where $g_X$ is the $U(1)_X$ gauge coupling constant and 
\eq{
\Lambda_L^{sb} = \frac{g_X \la_s \la^*_b v_\phi^2}{2 M_Q^2}.
}
This result is consistent with the one in~\cite{Altmannshofer:2014cfa,Sierra:2015fma} where diagrammatic method was used.
The $Z'$ boson does not couple directly to $\mu$ either, and the effective $Z'-\mu-\mu$ vertex is obtained by a similar procedure:
\eq{
 \Lambda_L^{\mu\mu}\, Z^{\prime \alpha}\, \ol{\mu}_L \gamma_\alpha \mu_L + h.c.,
\label{eq:Zprime-m-m}
}
where
\eq{
\quad \Lambda_L^{\mu\mu} = \frac{g_X |\la_\mu|^2 v_\phi^2}{2 M_L^2}.
}

\begin{figure}[tbp]
\begin{center}
\includegraphics[width=0.6\textwidth]{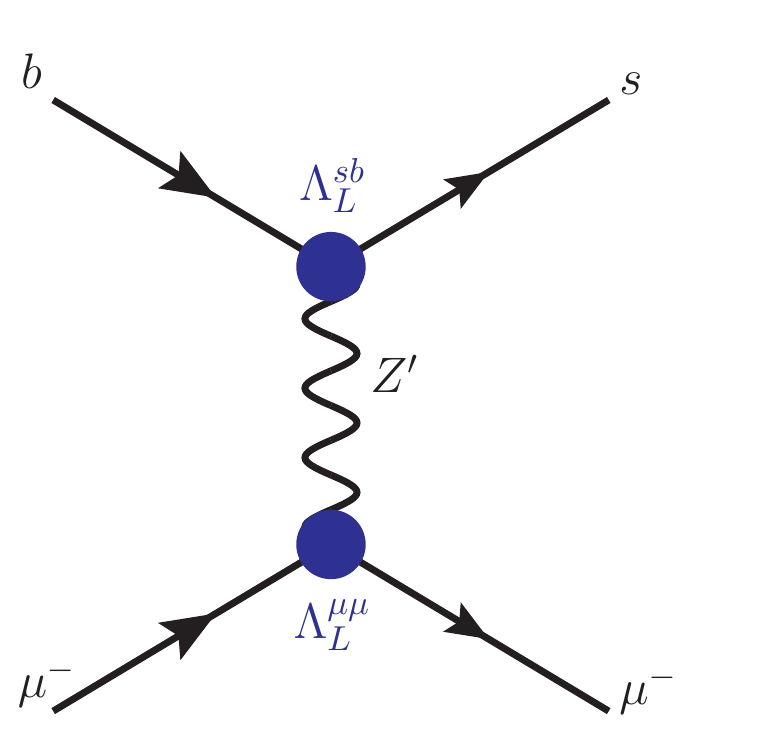}
\caption{Feynman diagram for $b \to s \mu^+\mu^-$.}
\label{fig:bsmm}
\end{center}
\end{figure}

The prediction of Wilson coefficients $C_9, C_{10}$ in the model can be made simply by integrating the mediating $Z'$ gauge boson out
in Fig.~\ref{fig:bsmm}:
\eq{
C_9 = -C_{10} = - \frac{\pi}{\sqrt{2} G_F \alpha_{\rm em}}\frac{1}{V_{ts}^* V_{tb}} \frac{\Lambda_L^{sb} \Lambda_L^{\mu\mu}}{m_{Z'}^2}.
\label{eq:C9_model}
}
By defining $x_i \equiv \la_i v_\phi/\sqrt{2} M_Q$, $x_\mu \equiv \la_\mu v_\phi/\sqrt{2} M_L$, and using the relation $m_{Z'} = g_X v_\phi$,
the above results can be rewritten as
\eq{
C_9 = -C_{10} &=-\frac{\pi}{\sqrt{2} G_F \alpha_{\rm em} V_{ts}^* V_{tb}} \frac{x_{sb} |x_\mu|^2}{v_\phi^2} \nl
&\approx -0.43 \,  \left(x_{sb} \over -0.001\right) \left|x_\mu \over 0.5\right|^2 \left(0.6 \, {\rm TeV} \over v_\phi\right)^2,
\label{eq:C9_num}
}
where $x_{sb} \equiv x_s x_b^*$.
We can explain (\ref{eq:C9_NP}) with relatively low $v_\phi (=0.6 \, {\rm TeV})$ when $x_{sb}=-0.001$ and $x_\mu =0.5$.

The effective vertex (\ref{eq:Zprime-s-b}) also generates  $B_s-\ol{B}_s$ mixing at tree-level, which turns out
the most stringent constraint in our model~\cite{Sierra:2015fma,DiLuzio:2017fdq}.
In our model the effective Hamiltonian for the mixing is
\eq{
{\cal H}_{\rm eff}^{\De B=2} =\frac{4 G_F}{\sqrt{2}} (V_{ts}^* V_{tb})^2 [ C_{sb}^{LL} (\ol{s}_L \ga_\mu b_L) (\ol{s}_L \ga^\mu b_L)+h.c.].
}
The Wilson coefficient $C_{sb}^{LL}$ is constrained by the mass difference $\Delta M_s$ of the $B_s$ system.
The experimental measurement~\cite{Amhis:2016xyh}
\eq{
\Delta M_s^{\rm exp} = (17.757 \pm 0.021) \, \text{ps}^{-1}
}
is smaller than the SM prediction\footnote{The prediction uses only tree-level inputs for the CKM parameters.}~\cite{DiLuzio:2017fdq}
\eq{
\Delta M_s^{\rm SM} = (19.9 \pm 1.5) \, \text{ps}^{-1}.
}
For the NP model which interferes with the SM constructively, as in our case when we take the coupling constants to be real, the constraint is very severe.
The NP contributes to $C_{sb}^{LL}$ through the $Z^\prime$-exchanging tree-level diagram with its effective
vertex in (\ref{eq:Zprime-s-b}):
\eq{
C_{sb}^{LL} = \frac{1}{4 \sqrt{2} G_F (V_{ts}^* V_{tb})^2}\frac{\left(\Lambda_L^{sb}\right)^2}{ m^2_{Z^\prime}}
= \frac{1}{4 \sqrt{2} G_F (V_{ts}^* V_{tb})^2}\frac{x_{sb}^2}{ v_\phi^2}.
\label{eq:C1_model}
}
Then we get 2$\si$ upper bound 
\eq{
\frac{x_{sb}}{v_\phi} \le 2.05 \times 10^{-3} \, \text{TeV}^{-1}.
\label{eq:xsb_constraint}
}

\begin{figure}[tbp]
\begin{center}
\includegraphics[width=0.6\textwidth]{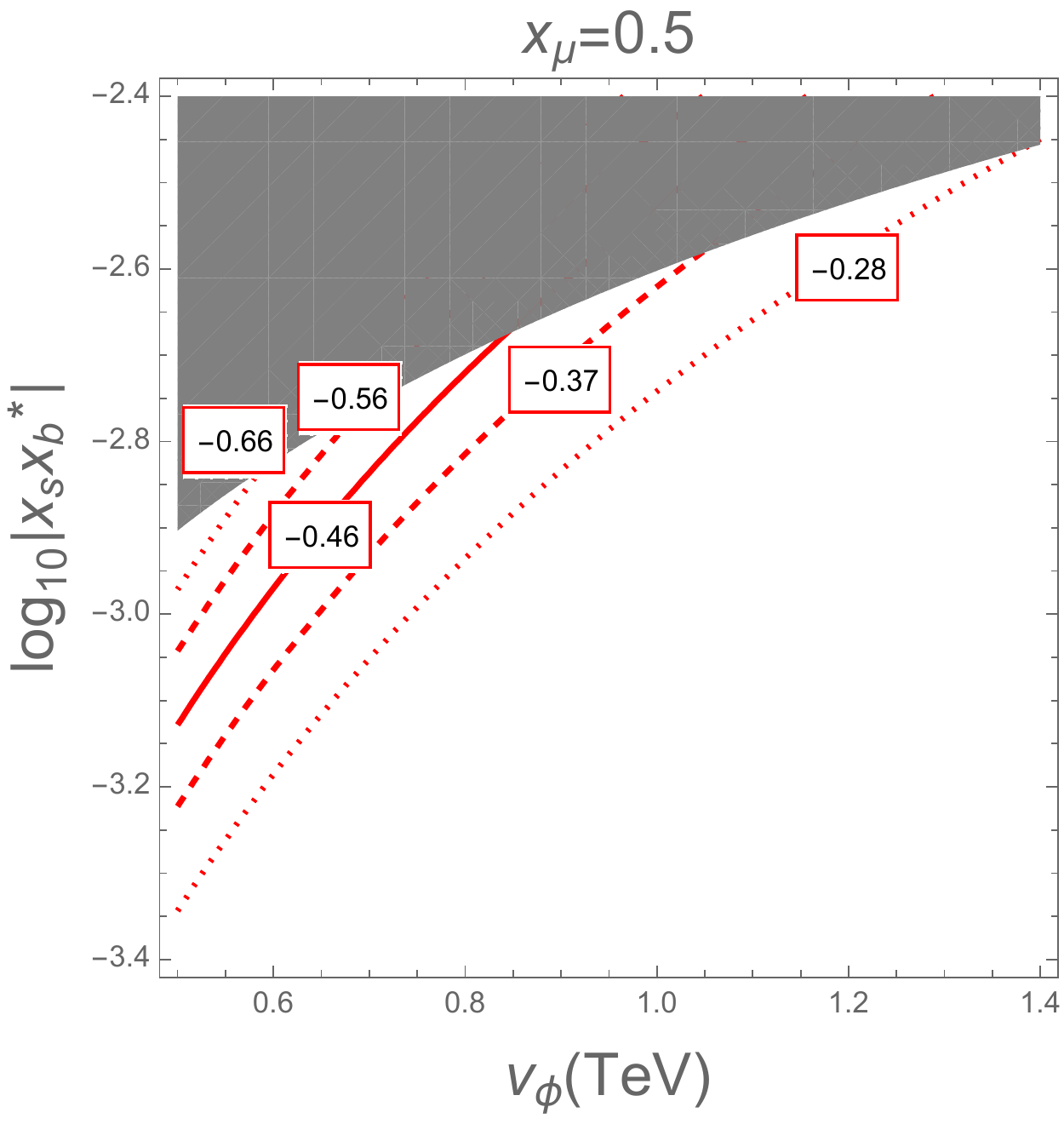}
\caption{The solid red line is the central value (\ref{eq:C9_NP}) to solve the $b \to s \mu^+ \mu^-$ puzzle.
The dashed (dotted) line represents 1$\si$ (2$\si$) region. 
The gray region is diafavoured
by the $B_s-\ol{B}_s$ mixing.  We fixed $x_\mu=0.5$.}
\label{fig:vphi-xsb}
\end{center}
\end{figure}

In Figure~\ref{fig:vphi-xsb} we show the region which can explain the $b \to s \mu^+ \mu^-$ puzzle in $(v_\phi, |x_{sb}|)$ plane.
The solid red line is the central value (\ref{eq:C9_NP}) to solve the $b \to s \mu^+ \mu^-$ puzzle.
The dashed (dotted) lines represent 1$\si$ (2$\si$) region. 
The gray region is diafavoured
by the $B_s-\ol{B}_s$ mixing.  We fixed $x_\mu=0.5$.


For $v_\phi =0.6$ TeV, $|x_{sb}|=10^{-3}$, and $|\la_s \la_b^*|=1$,  we get $M_Q \approx 13$ TeV.
This heavy quark mass is much smaller than the PQ-breaking scale, and therefore the Yukawa coupling should be very small $y_Q \sim 10^{-8}$.
Since  the symmetry is enhanced in the $y_Q \to 0$ limit\footnote{The PQ charges of $Q_{L,R}$ can be arbitrary.}, the small $y_Q$ is technically natural.

Since $Q_L(L_L)$  has the same quantum number with the SM $q_L (\ell_L)$ and $Q_R (L_R)$ does not mix with the SM $u_R, d_R (e_R, \nu_R)$,
the SM $Z$ boson coupling to the SM fermions are flavour-diagonal. Therefore the $Z$-mediated FCNC interactions are not
generated, which makes the constraint from the $Z$ boson interactions mild.
For other constraints for the model, we refer the reader to~\cite{Sierra:2015fma} where possible constraints are considered in detail.

\section{Conclusions}
\label{sec:conclusion}

There are several clues which suggest new physics beyond the standard model of particle physics: dark matter, neutrino mass,
strong CP problem, and possibly the flavour anomaly in $b \to s\mu^+ \mu^-$ transition.
We have considered a minimal model which can address the above hints simultaneously.

The neutrino mass is obtained when neutrino-specific Higgs-doublet gets a VEV, $v_1$~\cite{Baek:2016wml,Baek:2018wuo,Baek:2019wdn}. 
The value $v_1$  is due to the breaking of the PQ symmetry as shown in Figure~\ref{fig:neutrino_mass}, is
related to the high energy scale via seesaw-like relation (\ref{eq:m_nu}),
and is naturally small. Since the PQ symmetry does not allow the right-handed neutrino mass term, the neutrinos are Dirac-type.

The axion is KSVZ-type, but is distinguished from the original KSVZ model in that the axion-photon coupling is that of DFSZ-type axion model
and the axion-neutrino coupling is sizable.
The heavy quarks carrying PQ charges are also at 10 TeV scale without causing fine-tuning problem, and also can be tested in near future
experiments.

The heavy quarks and heavy leptons also carry the charges of $U(1)_X$ gauge symmetry.
The $U(1)_X$-gauge-boson-mediating tree-level diagram generates the quark-level $b \to s \mu^+ \mu^-$ transition, which
can explain the flavour puzzle.
We considered the constraint from the existing experiments including $B_s-\ol{B}_s$ mixing.

In the model, the neutrino mass and the axion are generated by the breaking the PQ-symmetry. The PQ-charged heavy quarks which
generate the QCD anomaly also induce the flavour-changing $Z^\prime$ couplings. The axion is also a good candidate for the cold dark matter.
Therefore our model shows an interplay between neutrino mass, flavour physics, axion, and dark matter.


\acknowledgments
This work was supported in part by the National Research Foundation of Korea(NRF) grant funded by
     the Korea government(MSIT), Grant No. NRF-2018R1A2A3075605.

\bibliographystyle{JHEP}
\bibliography{RK_axion}

\end{document}